\documentclass[10pt]{article}

\usepackage[authoryear]{natbib}

\usepackage[T1]{fontenc}
\usepackage[utf8x]{inputenc}
\usepackage[english,french]{babel}

\usepackage{graphicx}
\usepackage{hyperref}
\hypersetup{allcolors=black,colorlinks=true,breaklinks=true}
\usepackage{array}
\usepackage{enumitem}
\usepackage{booktabs}

\makeatletter
\newskip\@bigflushglue \@bigflushglue = -100pt plus 1fil

\def\bigcentering{\let\\\@centercr\rightskip\@bigflushglue%
\leftskip\@bigflushglue
\parindent\z@\parfillskip\z@skip}

\makeatother
\usepackage{aeguill}
\usepackage{times}

\title{Méthodologie pour identifier les terrains d'étude dans des corpus scientifiques}

\usepackage{authblk}

\author[1]{\'Eric Kergosien}
\author[2]{Marie-No\"elle Bessagnet}
\author[3]{Maguelonne Teisseire}
\author[1,5]{Joachim Sch\"opfel}
\author[4]{Amin Farvardin}
\author[1]{St\'ephane Chaudiron}
\author[1]{Bernard Jacquemin}
\author[2]{Annig Lacayrelle}
\author[3]{Mathieu Roche}
\author[2]{Christian Sallaberry}
\author[3]{Jean-Philippe Tonneau}

\affil[1]{Univ. Lille, EA 4073 \textendash{} GERiiCO, F-59000 Lille\protect\\prenom.nom @univ-lille.fr}
\affil[2]{LIUPPA, Universit\'e de Pau et des Pays de l'Adour, Pau\protect\\prenom.nom@univ-pau.fr}
\affil[3]{TETIS, Univ. Montpellier, APT, Cirad, CNRS, Irstea, Montpellier\protect\\prenom.nom @cirad.fr}
\affil[4]{LAMSADE, Universit\'e Paris-Dauphine, Paris\protect\\MohammadAmin.Farvardin@dauphine.eu}
\affil[5]{ANRT, Lille\protect\\prenom.nom @univ-lille.fr}

\date{\vspace{-2ex}}

\usepackage{textcomp}

\begin{document}

\maketitle

\begin{abstract}
\noindent Le projet interdisciplinaire TERRE-ISTEX a pour objectif d'identifier l'évolution des fronts de recherche en relation avec les territoires d'études, les croisements disciplinaires ainsi que les modalités concrètes de recherche à partir des contenus numériques hétérogènes disponibles dans les corpus scientifiques. Le projet se décompose en trois actions principales~: (1) identifier les périodes et les lieux qui ont fait l'objet d'études empiriques et dont rendent compte les publications issues des corpus analysés, (2) identifier les thématiques traitées dans le cadre de ces études et enfin (3) développer un démonstrateur Web de recherche d'information géographique (RIG). Les deux premières actions font intervenir des approches combinant des patrons du traitement automatique du langage naturel à des méthodes de fouille de textes. En croisant les trois dimensions (spatial, thématique et temporel) dans un moteur de RIG, il sera ainsi possible de comprendre quelles recherches ont été menées sur quels territoires et à quel moment. Dans le cadre du projet, les expérimentations sont menées sur un corpus hétérogène constitué de thèses électroniques et d'articles scientifiques provenant des bibliothèques numériques d'ISTEX et du centre de recherche CIRAD. \\[1ex] 
\textbf{Mots-clefs:} Fouille de textes, Traitement automatique des langues, Recherche d'information géographique, Scientométrie, Analyse du document.

\selectlanguage{english}
\begin{center}\textbf{Abstract}\end{center}
The TERRE-ISTEX project aims at identifying the evolution of research working relation to study areas, disciplinary crossings and concrete research methods based on the heterogeneous digital content available in scientific corpora. The project is divided into three main actions: (1) to identify the periods and places which have been the subject of empirical studies, and which reflect the publications resulting from the corpus analyzed, (2) to identify the thematics addressed in these works and (3) to develop a web-based geographical information retrieval tool (GIR). The first two actions involve approaches combining Natural languages processing patterns with text mining methods. By crossing the three dimensions (spatial, thematic and temporal) in a GIR engine, it will be possible to understand what research has been carried out on which territories and at what time. In the project, the experiments are carried out on a heterogeneous corpus including electronic thesis and scientific articles from the ISTEX digital libraries and the CIRAD research center. \\[1ex]
\selectlanguage{french}
\end{abstract}

\section*{Introduction}

L'accès quasi universel à des ressources
numériques, via des plateformes de bibliothèques \textendash{} par exemple, le
projet Gallica de la BnF\footnote{\url{http://gallica.bnf.fr/}}, Persée
ou la plateforme ISTEX\footnote{\url{http://www.istex.fr/}}, des
répertoires d'archives ouvertes (HAL-SHS), des
entrepôts de thèses électroniques (TEL), des services de
fédération de contenus (Isidore), ou encore des plateformes
d'édition électronique (tels que Cairn, ou encore
Revues.org) \textendash{} offre des opportunités d'usage
innombrables. Le projet ISTEX a pour objectif de créer des services
de recherche d'information innovants pour accéder à
l'ensemble de ces ressources numériques selon
différents critères. L'adoption croissante des
technologies de l'information et de la communication
par des disciplines, telles que les sciences humaines et sociales,
modifie les conditions d'appropriation des savoirs.
Ainsi, les humanités numériques ont permis de développer des
plateformes, mettant à disposition des chercheurs, des corpus et des
services d'aide à l'exploitation et
à la diffusion des dits corpus (par exemple,
l'application TELMA\footnote{\url{http://www.cn-telma.fr/}}).

Le projet TERRE-ISTEX\footnote{\url{http://terreistex.hypotheses.org/}}
s'inscrit dans cette mouvance et propose (1)
d'identifier les territoires étudiés dans les
contenus de corpus scientifiques disponibles en version numérique au
sein notamment de la bibliothèque ISTEX, et (2)
d'analyser les disciplines scientifiques impliquées
(histoire, géographie, sciences de l'information et
de la communication, sociologie, etc...) ainsi que
l'évolution des pratiques mono disciplinaires et/ou
pluridisciplinaires sur les territoires d'études
identifiés. L'intérêt de ces travaux est
notamment d'appuyer les scientifiques dans leur travail
de veille. Il est en effet primordial pour les chercheurs souhaitant
réaliser une étude scientifique sur un territoire (espaces non
urbain ou urbain à différentes échelles, tels que la commune, la
région, le pays ou même le continent) d'avoir
accès aux différentes études réalisées en amont sur ce même
territoire.

Nous pouvons citer d'autres projets ISTEX
complémentaires et d'un grand intérêt dans ces
recherches~: le projet
ALPAGE\footnote{\url{http://www.istex.fr/wp-content/uploads/2016/05/11.ISTEX_Chantiers_usage_ALPAGE.pdf}}
sur les aspects d'Annotation des corpus ISTEX et de
codage en TEI, et le projet LorExplor\footnote{\url{http://ticri.univ-lorraine.fr/wicri-musique.fr/index.php?title=Seminaire_ISTEX_2016}}
qui s'attaque aux résolutions de problèmes
éventuellement complexes menées dans un contexte de coopération
(accompagnement) entre les spécialistes du domaine
d'application et ceux du numérique.

Dans cet article, nous présentons la méthodologie mise en place pour
indexer automatiquement les corpus scientifiques afin
d'identifier les territoires étudiés. La section 2
présente les travaux connexes à l'extraction
d'entités nommées spatiales, temporelles et
thématiques pour l'identification
d'un territoire. Elle met également en avant le
manque d'outils pour la recherche
d'information géographique dans des corpus textuels
volumineux. La section 3 décrit la démarche générale utilisée
dans le projet, les jeux de données traitées, le modèle de
données proposé et l'étape importante de
normalisation des formats. La section 4 détaille les contributions
scientifiques pour l'extraction des informations
spatiales, thématiques et temporelles. La section 5 aborde la mise en
{\oe}uvre des expérimentations faites avec la chaîne de traitement
et les premiers résultats d'annotation obtenus. Elle
présente également l'application Web \textit{SISO}
pour appuyer les experts dans leur travail d'analyse et
de validation des corpus annotés. La section 6 conclut
l'article en évoquant les travaux en cours.

\section{Travaux connexes}

\subsection{La notion de Territoire}

Au-delà de sa stricte définition d'entité
administrative et politique, le territoire, selon Guy Di Méo
témoigne d'une «~appropriation à la
fois économique, idéologique et politique de
l'espace par des groupes qui se donnent une
représentation particulière d'eux-mêmes, de leur
histoire, de leur singularité~» \cite{di_meo_geographie_1998}.
Dans ce contexte éminemment subjectif, la caractérisation et la
compréhension des perceptions d'un même territoire
par les différents acteurs est difficile, mais néanmoins
particulièrement intéressante dans une perspective
d'aménagement du territoire \cite{derungs_text_2013}
et de politique publique territoriale. La notion de territoire fait
référence à différents concepts tels que les informations
spatiales et temporelles, les acteurs, les opinions,
l'histoire, la politique, etc. Dans cet article, nous
nous focalisons sur la détection d'entités
nommées (EN) de type lieu, que l'on nomme Entité
Spatiale (ES), d'entités thématiques et
d'entités temporelles.

\subsection{Extraction des entités nommées}

Les Entités Nommées (EN) ont été définies comme des noms de
personnes, des lieux et des organisations lors des campagnes
d'évaluations américaines appelées MUC (\textit{Message
Understanding Conferences}), qui furent organisées dans les années
90. Dans cet article, nous nous concentrons sur les lieux et les
entités temporelles et le premier défi consiste à reconnaître
ces types d'EN, le second étant
d'identifier les entités thématiques. 

De nombreuses méthodes permettent de reconnaître les EN en
général et les \textbf{entités spatiales} en particulier \cite{nadeau_survey_2007}. Parmi les méthodes d'extraction
d'informations s'appuyant sur des
textes, les approches statistiques étudient généralement les
termes co-occurrents par analyse de leur distribution dans un corpus
\cite{agirre_enriching_2000} ou par des mesures calculant la
probabilité d'occurrence d'un
ensemble de termes \cite{velardi_using_2001}. Ces méthodes ne
permettent pas toujours de qualifier des termes comme étant des EN,
notamment les EN de type ES. Des méthodes de fouille de données
fondées sur l'extraction de motifs permettent de
déterminer des règles (appelées règles de transduction) afin de
repérer les EN \cite{maurel_casen_2011}. Ces règles utilisent des
informations syntaxiques propres aux phrases \cite{maurel_casen_2011}. Des approches récentes s'appuient sur le Web
pour établir des liens entre des entités et leur type (ou
catégorie). Par exemple, l'approche de \cite{bonnefoy_lia-ismart_2011}
repose sur le principe que les distributions de
probabilités d'apparition des mots dans les pages
associées à une entité donnée sont proches des distributions
relatives aux types. Globalement, les relations peuvent être
identifiées par des calculs de similarité entre leurs contextes
syntaxiques \cite{grefenstette_explorations_1994}, par prédiction à
l'aide de réseaux bayésiens \cite{weissenbacher_identifier_2007}, par des techniques de fouille de textes \cite{grcar_using_2009}
ou encore par inférence de connaissances à
l'aide d'algorithmes
d'apprentissage \cite{giuliano_exploiting_2006}. Ces
méthodes sont efficaces, mais elles n'identifient pas
toujours la sémantique de la relation.

Pour la reconnaissance des classes d'EN, de nombreuses
approches s'appuient sur des méthodes
d'apprentissage supervisé. Ces méthodes
d'apprentissage comme les SVM \cite{joachims_text_1998} ou
encore les champs aléatoires conditionnels notés CRF \cite{mccallum_efficiently_2012,zidouni_structured_2009}
sont souvent utilisées dans le
challenge \textit{Conference on Natural Language Learning} (CoNLL). Les
algorithmes exploitent divers descripteurs ainsi que des données
expertisées/étiquetées. Les types de descripteurs utilisés sont
par exemple les positions des termes, les étiquettes grammaticales,
les informations lexicales (par exemple, majuscules/minuscules), les
affixes, l'ensemble des mots dans une fenêtre autour
du candidat, etc. \cite{carreras_simple_2003}. Dans
l'approche proposée dans cet article, nous combinons
de telles méthodes d'apprentissage supervisé
associées à des patrons linguistiques. Nous nous appuyons notamment
sur les travaux de \cite{lesbegueries_geographical_2006} pour la
définition de patrons linguistiques pour l'extraction
d'ES.

De nombreux travaux sont consacrés à
l'\textbf{analyse temporelle} du document. En accord
avec \cite{tapi_nzali_analyse_2015}, l'analyse des expressions
temporelles dans les textes est une problématique du traitement
automatique des langues qui connaît un intérêt grandissant depuis
quelques années. Les travaux de recherche montrent la diversité
dans la langue des documents tels que des textes journalistiques en
langue anglaise ou encore des textes dans d'autres
langues (chinois, français, suédois, {\dots}) 
\cite{li_chinese_2014,parc-lacayrelle_composante_2007,strotgen_time_2014,moriceau_french_2013}.
Les travaux dans ce domaine
montrent également la diversité dans le type de document traité
tels que les SMS, les textes historiques, les résumés
d'essais cliniques, des monographies relatant des
récits de voyages, ou encore des articles scientifiques.
L'annotation temporelle permet
d'extraire et de normaliser des expressions temporelles
qui pourront ensuite être utilisées dans un contexte de recherche
d'information géographique combinant les dimensions
spatiale, thématique et temporelle. Normaliser revient à
transformer une expression (par exemple,
«~avant-hier~» ou «~le
1\textsuperscript{er} janvier 2017~») en une représentation formatée
et spécifiée. Plusieurs outils mettent en {\oe}uvre une telle
annotation temporelle. On peut citer SUTime \cite{chang_sutime_2012}
pour la langue anglaise, XIP \cite{bittar_annotateur_2012} et HeidelTime
\cite{strotgen_multilingual_2013} pour des adaptations multilingues. Au
regard des résultats de ces outils sur les langues française et
anglaise, nous intégrerons HeidelTime dans notre chaîne de
traitement.

Afin de compléter les connaissances identifiées, nous souhaitons
également mettre en perspective des modules de fouille de textes pour
extraire \textbf{les} \textbf{thématiques}. Dans ce sens, les termes
constituent en effet la base de ressources sémantiques ou thésaurus
du domaine général \cite{kennedy_automatically_2010,vakkari_how_2010} ou
spécialisé \cite{turenne_beluga_2004,bartol_assessment_2009,neveol_language_2014}.
Leur construction peut être guidée (1) par
consensus avec les experts \cite{laporte_thesauformtraits_2012}, (2) par les
données nécessitant, par exemple, la mise en {\oe}uvre de
méthodes de Fouille de Textes \cite{dobrov_combining_2011}. Notre
travail se positionne sur ce deuxième point. Les méthodes
classiques d'extraction de la terminologie sont
fondées sur des approches statistiques et/ou syntaxiques. Le
système \textit{TERMINO} \cite{david_necessite_1990} est un outil
précurseur qui s'appuie sur une analyse morphologique
à base de règles pour extraire les termes nominaux (aussi appelés
syntagmes nominaux). Les travaux de \cite{smadja_retrieving_1993} (approche
\textit{XTRACT}) s'appuient sur une approche
statistique. \textit{XTRACT} extrait, dans un premier temps, les
syntagmes binaires situés dans une fenêtre de dix mots. Les
syntagmes binaires sélectionnés sont ceux qui dépassent
d'une manière statistiquement significative la
fréquence due au hasard. L'étape suivante consiste
à extraire les groupes de mots contenant les syntagmes binaires
trouvés à la précédente étape. \textit{ACABIT} \cite{daille_approche_1994}
effectue une analyse linguistique afin de transformer les syntagmes
nominaux en termes binaires. Ces derniers sont ensuite triés selon
des mesures d'association entre éléments composant
les syntagmes. Les mesures d'association et les
approches distributionnelles ont été étendues et adaptées pour
extraire des termes spécialisés \cite{frantzi_automatic_2000} et
identifier des termes synonymes \cite{daille_semi-compositional_2014}. Contrairement
à ACABIT qui est essentiellement fondé sur des méthodes
statistiques, LEXTER et SYNTEX s'appuient, en grande
partie, sur une analyse syntaxique approfondie \cite{bourigault_approche_2000}.
La méthode consiste à extraire les syntagmes nominaux
maximaux. Ces derniers sont alors décomposés en termes de
«~têtes~» et d'«~expansions~» à
l'aide de règles grammaticales. Les termes sont alors
proposés sous forme de réseau organisé en fonction de critères
syntaxiques. Des plateformes qui s'appuient sur les
caractéristiques principales de cet état de l'art
ont été implantées. Dans ce contexte, nous pouvons citer
TermSuite proposé par le LINA \cite{daille_ttc_2011} qui suit
une méthodologie en 4 phases~: (1) Pré-traitements~; (2) Analyses
linguistiques (découpage du texte en mots, analyse morphosyntaxique
et lemmatisation)~; (3) Extraction terminologique monolingue (termes
simples et complexes)~; (4) Alignement terminologique bilingue. Une
seconde plateforme nommée BioTex, développée par
l'équipe TETIS de Montpellier \cite{lossio-ventura_biomedical_2016}
exploite à la fois des informations
statistiques et linguistiques pour extraire la terminologie à partir
de textes libres. BioTex s'appuie sur une
méthodologie générique qui a été essentiellement appliquée
aux domaines scientifiques biomédical et agronomique. Dans le cadre
du projet TERRE-ISTEX, nous souhaitons mettre en place une approche
classique de marquage sémantique en nous appuyant sur le thésaurus
Agrovoc\footnote{\url{http://aims.fao.org/fr/agrovoc}}. Nous souhaitons à
terme étendre notre approche en adaptant la chaîne de traitement
développée dans l'application BioTex, à
l'ensemble des articles collectés sur la
thématique Changement Climatique, et plus généralement aux
différentes disciplines scientifiques. 

\subsection{Contexte pour l'analyse géographique de la littérature scientifique}

La scientométrie se réfère à l'étude de tous
les aspects de la littérature liée aux sciences et à la
technologie \cite{hood_literature_2001}. Cela implique des analyses
quantitatives sur les activités scientifiques, notamment les
publications. Ainsi, on tente d'apprécier divers
critères tels que l'évolution des pratiques des
chercheurs, le rôle des sciences et de la technologie sur les
économies nationales, l'évolution des technologies,
etc. Il existe aujourd'hui des sources
d'information publiques sur les publications
scientifiques telles que Google Scholar, les archives ouvertes (HAL par
exemple) permettant notamment d'analyser la production
des chercheurs. Nous pouvons également accéder à des bases de
données bibliographiques (SCOPUS, AERES, ...) pour enrichir les
analyses. Dans le domaine de l'informatique, des
analyses scientométriques ont été menées pour connaître
l'évolution des pratiques des chercheurs concernant
les articles publiés sur un corpus tel que la base de données DBLP
\cite{cavero_computer_2014} ou encore évaluer les collaborations
des chercheurs dans le cadre de leurs publications \cite{cabanac_academic_2015}.
Dans le cadre de la campagne d'acquisition
et d'indexation des archives scientifiques pour créer
la bibliothèque numérique ISTEX, il est indispensable de proposer
des services de recherche d'information innovants pour
accéder à l'ensemble de ces ressources numériques
selon différents critères, et notamment les informations spatiales,
temporelles et thématiques présentes dans les documents.

La RI géographique (RIG), nommée et définie pour la première
fois par Ray Larson \cite{larson_geographic_1995}, est la tâche de recherche de
documents satisfaisant des caractéristiques géographiques, la
notion de géographie associant de façon explicite la dimension
temporelle à la dimension spatiale et/ou thématique \cite{martins_geographically-aware_2007,liu_spatio-temporal-textual_2010}. À notre
connaissance, de tels travaux de RIG n'ont
jusqu'à présent pas été associés à ceux de
scientométrie. Cependant, la combinaison ces trois dimensions
(spatiale, temporelle et thématique) semble importante pour
améliorer les analyses.

Un récent travail intitulé «~Anthologie des congrès
Inforsid~»\footnote{\url{http://dbrech.irit.fr/rechpub/cabanac_inforsid.accueil/}} présente
une analyse des 30 éditions du congrès. Il liste les thèmes de la
conférence au fil des années et les villes des conférences et des
laboratoires d'affiliation des auteurs. Enfin, une base
de données permet d'accéder aux informations
relatives aux articles et aux auteurs. Au delà des analyses de ce
premier travail, dédiées à la série de conférences Inforsid,
nous proposons une approche générique applicable aux données
relatives à tout corpus constitué d'articles
scientifiques. Notre approche supporte des opérations
d'extraction d'informations combinant
les dimensions spatiale, temporelle et thématique. La démarche
générale ainsi que le modèle de données résultant de la phase
d'indexation sont illustrés dans la section suivante.

\section{Le projet TERRE-ISTEX}

\subsection{Démarche générale}

\begin{figure}[!h]
 \includegraphics[width=.85\textwidth]{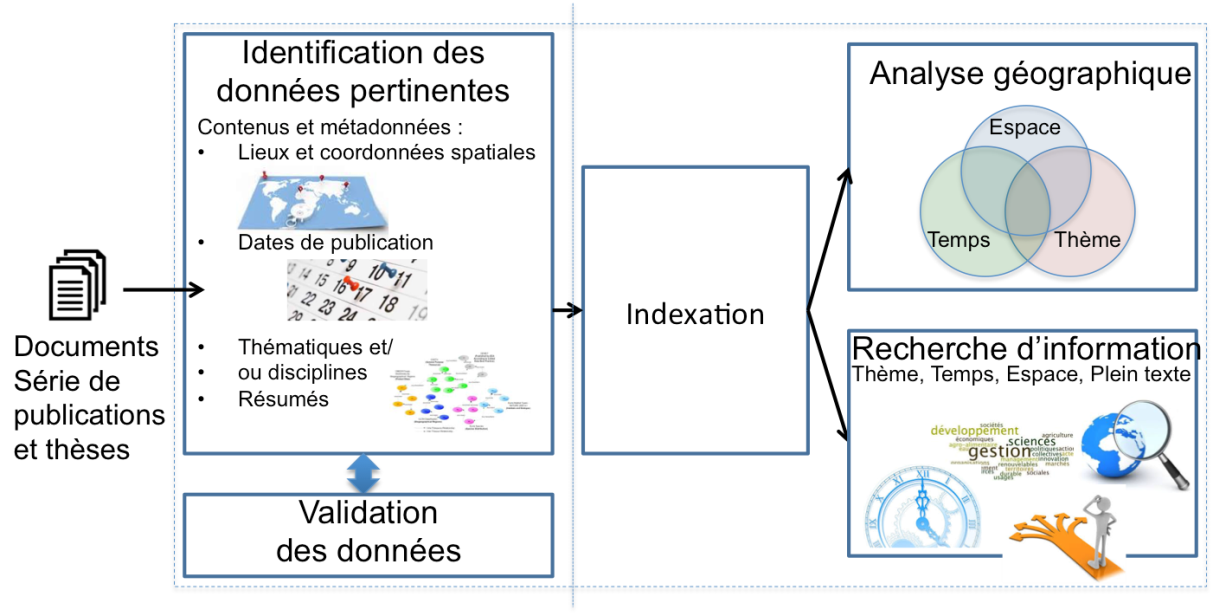}
 \caption{Démarche générique pour l'analyse multidimensionnelle de corpus de publications scientifiques.}
 \label{fig:demarcheGenerique}
\end{figure}

La méthodologie générique mise en {\oe}uvre dans le projet
TERRE-ISTEX est décrite par la Figure~\ref{fig:demarcheGenerique} \cite{kergosien_using_2017}.
Indépendamment de tout corpus de publications scientifiques, une
première étape vise à normaliser les corpus documentaires. Une
seconde étape vise à identifier, dans les métadonnées et les
contenus des documents, les terrains d'études ainsi
que les disciplines scientifiques impliquées. Nous entendons par
terrain d'études le ou les lieu(x) constituant le
territoire sur lequel est menée l'étude à une
date ou une période donnée.

Présentons le corpus traité dans le cadre de notre projet.

\subsection{Le corpus}

La constitution d'un corpus est
une étape préalable majeure dans un processus
d'analyse et de recherche d'information
comme celui que nous décrivons. Aussi, nous avons ciblé trois
sources d'information contenant des publications
scientifiques, à savoir les plateformes ISTEX\footnote{\url{http://www.istex.fr/category/plateforme/}} et
Agritrop\footnote{\url{https://agritrop.cirad.fr/}} (archive
ouverte du CIRAD\footnote{\url{http://www.cirad.fr}}), ainsi que les thèses de
l'ANRT\footnote{\url{https://anrt.univ-lille3.fr/}} et des
métadonnées associées disponibles sur le portail
theses.fr\footnote{\url{http://www.theses.fr/}}.

Le cas d'étude permettant de
mettre en pratique nos approches est le changement climatique sur les
territoires du Sénégal et de Madagascar. À cette fin, nous avons
collecté un corpus initial de documents provenant de la plateforme
ISTEX (environ 170\,000 documents) à partir de requêtes avec les
mots-clés suivants~: «~climate
change~», «~changement climatique~», «~Senegal~», «~Sénégal~»,
«~Madagascar~». À partir de ce même
ensemble de mots-clés, nous avons collecté 400 thèses provenant
de l'ANRT. Enfin, les documents provenant
d'Agritrop ciblent des études traitant de Madagascar
et du fleuve Sénégal. Sur les 92\,000 références et 25\,000
documents en texte intégral, nous pouvons
recenser différents genres~: des publications scientifiques et de la
littérature grise (des rapports, etc.). Chaque
document possède, en plus de son contenu, des métadonnées et un
résumé.

Selon la provenance du document, les
métadonnées sont soit au format MODS\footnote{\url{http://www.bnf.fr/fr/professionnels/f_mods/s.mods_presentation.html}}
(ISTEX), soit dans un format XML inspiré du Dublin Core (CIRAD), soit
en RDF (thèses ANRT). Le corpus est multilingue~: certains documents
sont en français et d'autres en anglais, mais nous
pouvons également trouver des documents utilisant les deux langues
(par exemple, ils comportent un résumé en français et un
résumé en anglais). Nous sommes ainsi confrontés à un ensemble
de documents multilingues et hétérogènes, à la fois dans leur
contenu mais également dans leur format.

\subsection{Une étape primordiale de transformation et de normalisation}

\subsubsection{La chaîne de traitement TERRE-ISTEX}

\begin{figure}[!h]
 \includegraphics[width=.85\textwidth]{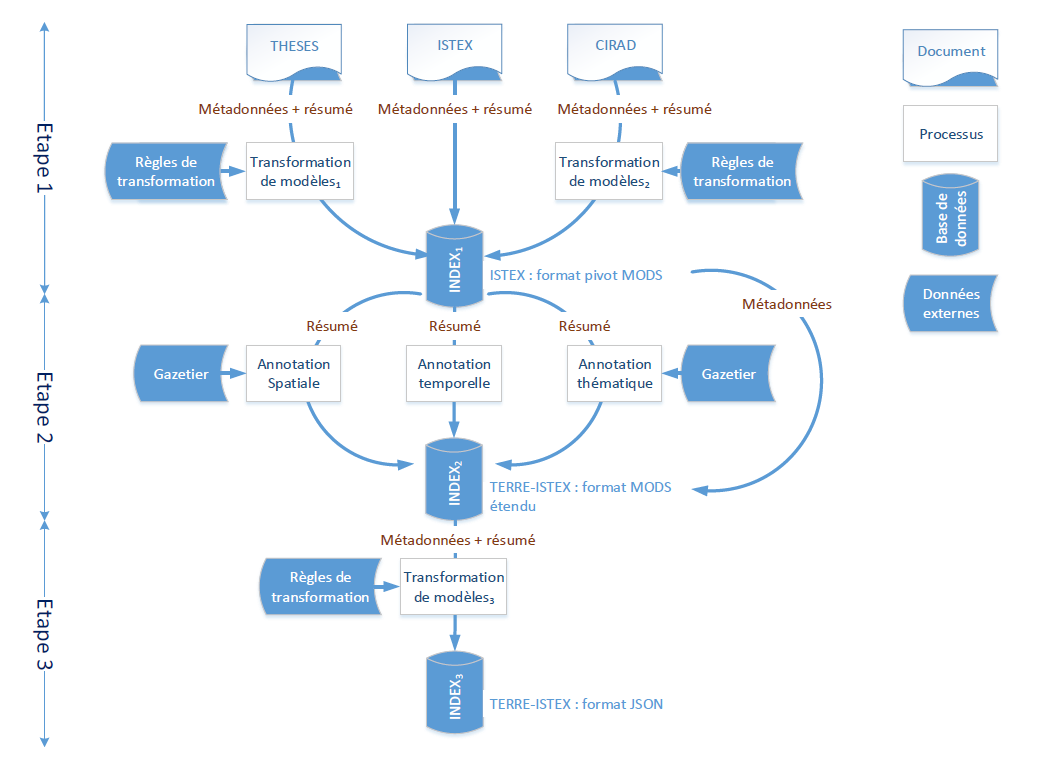}
 \caption{La chaîne de traitement TERRE-ISTEX pour l'identification des terrains d'études dans les corpus scientifiques.}
 \label{fig:chaineTerreIstex}
\end{figure}

La Figure~\ref{fig:chaineTerreIstex} décrit la chaîne de traitement
développée dans le projet TERRE-ISTEX. Dans
un premier temps, cette chaîne n'est appliquée que
sur les méta-données et les résumés. Pour pallier
l'hétérogénéité des données, nous avons
choisi de normaliser les métadonnées en utilisant le format pivot
MODS (\textit{Metadata Object Description Schema}), préconisé sur la
plateforme ISTEX. Le format MODS a plusieurs atouts~: (a) il est
approprié pour la description de tout type de document et tout
support (numérique ou non)~; (b) il est plus riche que le Dublin Core~;
(c) il est proche des modèles de
structuration des informations bibliographiques utilisés dans les
bibliothèques (par exemple, le format MARC). Ainsi, nous appliquons
un premier algorithme de transformation de modèle sur les 92\,400
documents de notre corpus ne respectant pas ce format (Étape 1, Figure~\ref{fig:chaineTerreIstex}). L'étape 2 concerne l'annotation,
dans les résumés, des entités spatiales, temporelles et
thématiques. Cette étape est détaillée
en section 4. En résultat, le modèle de
données MODS-TI étend le format MODS afin de décrire les
entités spatiales, temporelles et thématiques extraites des
documents. Le modèle MODS-TI est détaillé dans la section
suivante. L'étape 3 met en {\oe}uvre un nouvel
algorithme de transformation du format MODS-TI afin de créer les
index pour que l'ensemble des données puisse ensuite
être traité dans les dernières phases d'analyse
et de recherche d'information.

\subsubsection{Le modèle de données}

Le modèle de données MODS-TI étend donc le
format MODS afin de lui permettre de décrire les informations
spatiales, temporelles et thématiques extraites des documents et de
leurs méta-données.

Ainsi, nous avons ajouté trois balises à un
document MODS~:
\begin{itemize}
 \item \texttt{<spatialAnnotations>},
 \item \texttt{<temporalAnnotations>},
 \item \texttt{<thematicAnnotations>}.
\end{itemize}

\begin{figure}[!h]
 \includegraphics[width=.85\textwidth]{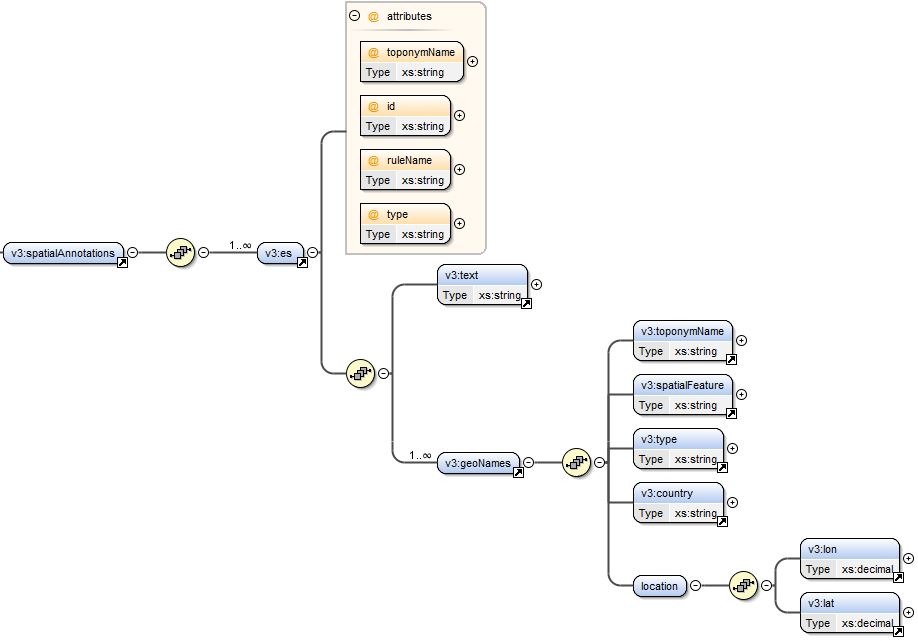}
 \caption{DTD décrivant la balise \texttt{<spatialAnnotations>}.}
 \label{fig:dtdSpatialAnnotations}
\end{figure}

La balise \texttt{<spatialAnnotations>}
contient un ensemble d'entités spatiales (balise
\texttt{<es>}), avec pour chacune d'elle, le texte annoté (balise
\texttt{<text>}) ainsi que son empreinte spatiale obtenue en interrogeant la ressource
Geonames. La DTD correspondante est donnée Figure~\ref{fig:dtdSpatialAnnotations}.

\begin{figure}[!h]
 \includegraphics[width=.85\textwidth]{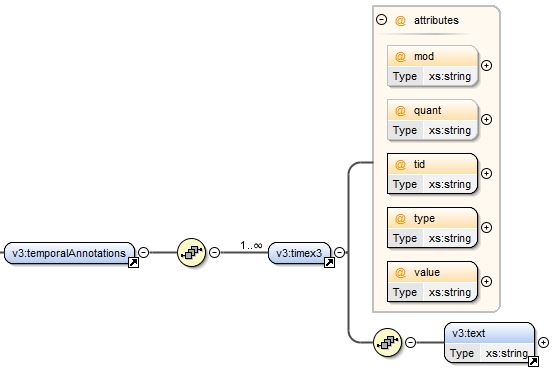}
 \caption{DTD décrivant la balise \texttt{<temporalAnnotations>}.}
 \label{fig:dtdTemporalAnnotations}
\end{figure}

La balise \texttt{<temporalAnnotations>}
contient un ensemble d'entités temporelles décrites
par les balises \texttt{<timex3>}
provenant d'Heildeltime complété par le texte
annoté (balise \texttt{<text>}).
La DTD correspondante est donnée Figure~\ref{fig:dtdTemporalAnnotations}.

\begin{figure}[!h]
 \includegraphics[width=.85\textwidth]{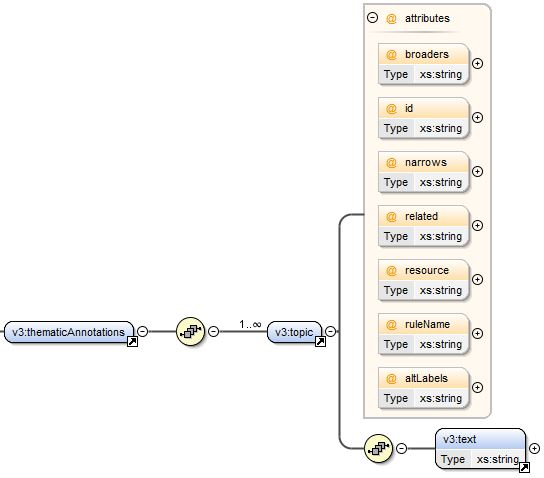}
 \caption{DTD décrivant la balise \texttt{<thematicAnnotations>}.}
 \label{fig:dtdThematicAnnotations}
\end{figure}

Enfin, la balise \texttt{<thematicAnnotations>}
contient l'ensemble des thèmes abordés dans le
résumé (balise \texttt{<topic>}),
avec pour chacun d'eux des informations provenant de la
ressource Agrovoc, en complément du texte annoté (balise
\texttt{<text>}). La DTD correspondante est donnée Figure~\ref{fig:dtdThematicAnnotations}.

\section{Annotation des entités}

\subsection{Les entités spatiales}

Dans l'approche proposée dans ce projet, nous
utilisons une méthodologie basée sur des patrons linguistiques pour
l'identification automatique des entités spatiales
(ES) \cite{tahrat_text2geo_2013}. Dans nos travaux, une entité
spatiale ES est composée d'au moins une entité
nommée et un ou plusieurs indicateurs spatiaux spécifiant son
emplacement. Une ES peut alors être identifiée de deux façons
\cite{sallaberry_semantic_2009}~: une ES absolue (ESA) est une
référence directe à un espace géo-localisable (par exemple
«~le plateau
d'Allada~»)~; une ES relative (ESR)
est définie à l'aide d'au moins une
ESA et d'indicateurs spatiaux d'ordre
topologique (par exemple, «~au
sud du Bénin~»). Ces indicateurs spatiaux
représentent des relations et nous en considérons cinq types dans
ces travaux~: l'orientation, la distance,
l'adjacence, l'inclusion et la figure
géométrique qui définit l'union ou
l'intersection liant au moins deux ES. Un exemple de ce
type d'ES est «~Near Paris~». 

Notons que l'avantage de notre représentation
intégrant les ESA et ESR réduit
significativement les ambiguïtés liées à
l'identification de la bonne empreinte spatiale. En
effet, le fait de prendre en compte les indicateurs spatiaux (par
exemple «~fleuve~» pour
«~fleuve Sénégal~») nous permet
d'identifier dans GEoNames la bonne empreinte spatiale.
Pour ce qui est des éventuels cas o\`u plusieurs entités spatiales
distinctes portant le même nom seraient identifiées dans le corpus
(exemple Bayonne en France et Bayonne aux États-Unis), nous analysons
le contexte dans le document textuel traité pour proposer une
désambiguïsation \cite{kergosien_when_2015}.

Afin d'identifier les entités spatiales, nous
appliquons et étendons un processus de TALN adapté au domaine de la
Recherche d'Information Géographique (RIG). Dans ce
contexte, des règles (patrons) des travaux de ont été
améliorés et intégrés pour identifier les entités spatiales
absolues et relatives dans le corpus en français (e.g.
\textit{sud-ouest de l'Arabie Saoudite} (ESR),
\textit{dans la région du Mackenzie} (ESR),~\textit{golfe de
Guinée} (ESA), \textit{lac Eyre} (ESA)), et dans le corpus en anglais
(e.g. \textit{Willamette River} (ESA), \textit{Indian Ocean} (ESA),
\textit{Wujiang River Basin} (ESA)) du projet TERRE-ISTEX. De plus,
nous proposons de nouvelles règles pour identifier les Organisations
(par exemple, «~une Organisation est suivie par un
verbe d'action~»). Ces différentes
règles ont été développées dans
l'environnement Gate permettant de désambiguïser les
entités extraites.

\subsection{Les entités thématiques et temporelles}

Afin de compléter les connaissances identifiées dans les
métadonnées et de préciser les sous domaines étudiés, nous
souhaitons appliquer sur le contenu des publications des modules en
fouille de textes pour l'extraction de vocabulaires de
domaine, Dans un premier temps, nous utilisons des ressources
sémantiques de domaine pour une annotation lexicale. Les entités
thématiques à annoter étant liées, dans notre cas, au domaine
du changement climatique, nous nous appuyons sur la ressource Agrovoc
\cite{rajbhandari_agrovoc_2012}. Cette dernière est formalisée en XML
SKOS. Dans la phase d'indexation, nous marquons pour
chaque terme du contenu d'un article provenant
d'Agrovoc les termes «~employé
pour~» et les termes génériques, informations qui
seront exploitées dans le moteur de recherche. Nous visons à terme
à proposer une approche générique en donnant la possibilité
d'intégrer aisément une nouvelle ressource
sémantique de domaine formalisée en XML SKOS. Aussi, nous projetons
d'intégrer le module Biotex développé par
l'équipe TETIS de Montpellier \cite{lossio-ventura_biomedical_2016}
combinant des approches statistiques et
linguistiques pour extraire la terminologie à partir de textes
libres. Les informations statistiques apportent une pondération des
termes candidats extraits. Cependant, la fréquence
d'un terme n'est pas nécessairement
un critère de sélection adapté. Dans ce contexte, Biotex propose
de mesurer l'association entre les mots composant un
terme en utilisant une mesure appelée C-value tout en intégrant
différentes pondérations (TF-IDF, Okapi). Le but de C-value est
d'améliorer l'extraction des %
%On parle aussi de multi{}-terme
%
%Qu'est ce qu'un terme complexe
termes complexes (termes constitués de plusieurs mots),
particulièrement adaptés pour les domaines de spécialité.

En ce qui concerne les entités temporelles, nous avons intégré la
chaîne de traitement HeidelTime \cite{strotgen_multilingual_2013} permettant
de marquer des entités calendaires (dates et périodes). HeidelTime
est un système libre, à base de règles,
d'étiquetage d'expressions
temporelles, décliné dans plusieurs langues. Concernant
l'anglais, plusieurs corpus de documents (articles
scientifiques, presse) ont été traités \cite{strotgen_multilingual_2013}.
L'évaluation de ce système montre de
meilleurs résultats pour l'extraction et la
normalisation des expressions temporelles pour
l'anglais, dans le contexte des campagnes TempEval-2 et
TempEval-3 \cite{uzzaman_semeval-2013_2013} et étendu à 11 langues
dont le français \cite{moriceau_french_2013}. HeidelTime produit
des annotations dans le format ISO-TimeML, qui fait la distinction
entre quatre catégories d'expressions temporelles~:
les dates, les heures, les durées et les fréquences. Notre objectif
étant de connaître les périodes abordées dans les documents,
nous nous intéressons seulement aux expressions temporelles à %
%MNB, ce sont les dates et périodes 
%
%Est{}-ce que cela ne concerne que les dates
connotation calendaire (dates et périodes).

\section{Expérimentations}

\subsection{Premières expérimentations}

Les différents experts géographes et sociologues participant au
projet ont évalué les services d'extraction des
descripteurs ES en utilisant \textit{SISO}. Pour initier
l'évaluation en cours de projet, un premier corpus
d'articles de presse en langue française composé
de 4\,328 mots (71 ES et 117 Organisations) a été utilisé. Les
évaluations s'appuyant sur des mesures classiques
(précision, rappel et F-mesure) ont été menées en comparant les
résultats obtenus par une extraction manuelle réalisée par les
experts avec les résultats issus de la chaîne de traitement.
Concernant les ES, nous obtenons un excellent rappel (0.91), une
précision correcte (0.62), la valeur de F-mesure étant alors de
0.74. La grande majorité des ES est ainsi extraite (ceci est
illustré par le rappel élevé) mais les règles de marquage
engendrent encore différentes erreurs d'o\`u une
précision plus fiable. 

Dans un second temps, afin d'évaluer notre approche
d'annotation des ESA et ESR sur un corpus scientifique,
nous avons annoté manuellement 10 articles scientifiques en
français et 10 en anglais à partir du corpus collecté sur le
thème changement climatique. Les articles sont sélectionnés
aléatoirement. Les documents font en moyenne 230 mots et contenaient
39 entités spatiales (ESA, ESR). Nous avons ensuite annoté ces
documents avec deux chaînes de traitement~: la nôtre et la chaîne
CASEN, référence dans le domaine pour le marquage des entités
nommées \cite{maurel_casen_2011}. 

Nous avons obtenu de très bons résultats en termes de précision,
rappel et F-mesure avec notre chaîne de traitement (cf. Tableaux~\ref{tab:EvalFrancais} et \ref{tab:EvalAnglais}).

\begin{table}[!h]
 \centering
 \begin{tabular}{lcc}
 \hline
  & ESA, ESR & ESA, ESR \\
  & (TERRE-ISTEX) & (CASEN) \\
 \hline
  Précision & 100\% & 93\% \\
  Rappel & 90\% & 77\% \\
  F-mesure & 94,7\% & 84,2\% \\
 \hline
 \end{tabular}
 \caption{Évaluation du marquage des entités spatiales sur 10 articles en français.}
 \label{tab:EvalFrancais}
%  $\begin{array}{l|cc}
%   & ESA, ESR & ESA, ESR \\
%   & (TERRE-ISTEX) & (CASEN) \\
%  \hline
%   Pr\acute{e}cision & 100\% & 93\% \\
%   Rappel & 90\% & 77\% \\
%   F-mesure & 94,7\% & 84,2\% \\
%  \end{array}$
\end{table}

\begin{table}[!h]
 \centering
 \begin{tabular}{lcc}
 \hline
  & ESA, ESR & ESA, ESR \\
  & (TERRE-ISTEX) & (CASEN) \\
 \hline
  Précision & 90\% & 94\% \\
  Rappel & 60\% & 53,3\% \\
  F-mesure & 72\% & 68\% \\
 \hline
 \end{tabular}
 \caption{Évaluation du marquage des entités spatiales sur 10 articles en anglais.}
 \label{tab:EvalAnglais}
%  $\begin{array}{l|cc}
%   & ESA, ESR & ESA, ESR \\
%   & (TERRE-ISTEX) & (CASEN) \\
%  \hline
%   Pr\acute{e}cision & 90\% & 94\% \\
%   Rappel & 60\% & 53,3\% \\
%   F-mesure & 72\% & 68\% \\
%  \end{array}$
\end{table}

Le problème de la production et la mise à disposition
d'un corpus annoté (intégrant notamment les
entités spatiales, temporelles et thématiques) est également
partagé par d'autres équipes de projets ISTEX. À
ce sujet, une réflexion est actuellement en cours au sein de
l'équipe ISTEX pour travailler à la production
d'un corpus d'évaluation pour la
communauté, et dans notre cas, pour mener une évaluation à plus
grande échelle.

Pour appuyer le travail d'experts dans
l'annotation d'entités dans des
corpus, nous proposons l'application Web
\textit{SISO}\footnote{\url{http://geriico-demo.univ-lille3.fr/siso/}}
que nous détaillons dans la section suivante. 

\subsection{L'application \emph{SISO} pour l'aide à l'indexation de corpus}

\begin{figure}[!h]
 \includegraphics[width=.85\textwidth]{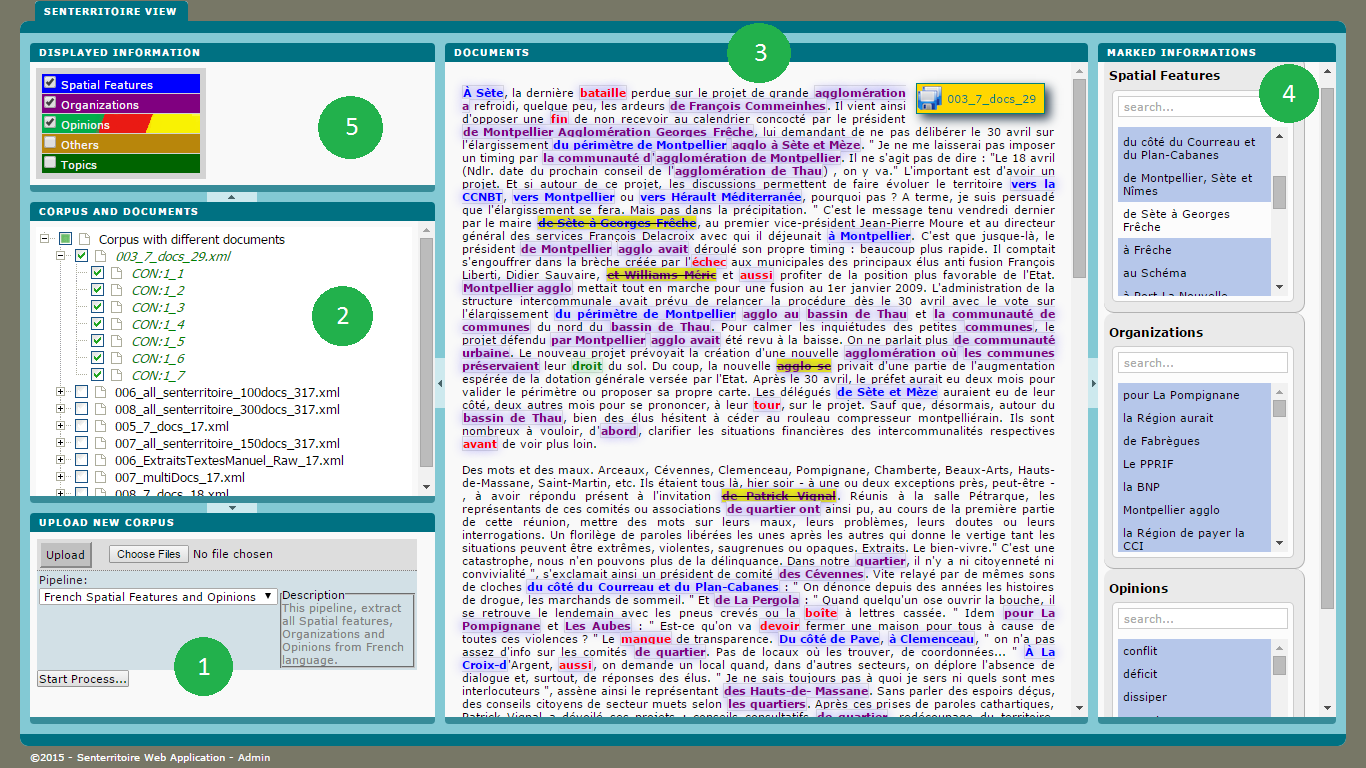}
 \caption{Le système Web \emph{SISO}.}
 \label{fig:SISO}
\end{figure}

L'application \textit{SISO} (Figure~\ref{fig:SISO}) permet aux
utilisateurs de télécharger des corpus, de les indexer pour marquer
différents types d'information, notamment les
entités spatiales, temporelles et thématiques. Les chaînes de
traitement développées pour l'extraction des
descripteurs s'appuient sur le système
Gate\footnote{\url{https://gate.ac.uk}}. L'application
permet ensuite de visualiser et de corriger manuellement les
résultats, puis d'exporter les résultats validés
au format XML. 

Via l'interface Web (Figure~\ref{fig:SISO}),
l'utilisateur expert peut télécharger et indexer un
corpus (\textit{frame} 1), pour pouvoir ensuite l'analyser
(\textit{frame} 2). Une fois les documents téléchargés et indexés,
l'utilisateur expert peut sélectionner les
descripteurs marqués qu'il souhaite visualiser (\textit{frame}
5), puis analyser les résultats (\textit{frame} 3) sur les documents
préalablement sélectionnés (\textit{frame} 2). Après avoir
sélectionné les types de descripteurs à afficher dans la \textit{frame} 5,
les informations correspondantes sont colorées dans les textes (\textit{frame}
3) et listées par catégorie (\textit{frame} 4). 

Dans le cas o\`u une erreur d'indexation est
identifiée par l'utilisateur expert, le marquage
réalisé peut être supprimé en enlevant le descripteur
concerné dans les listes présentées (\textit{frame} 4). 

L'utilisateur expert a également la possibilité
d'exporter et de récupérer au format XML le corpus
qu'il a sélectionné, analysé et éventuellement
corrigé. Dans un souci d'autonomie dans
l'usage de cette application, des utilisateurs avertis
peuvent intégrer et éditer leurs propres chaînes
d'indexation définies selon le format
Gate. Les corpus traités, les chaînes
d'indexation existantes ainsi que les lexiques peuvent
être mis à jour sur le serveur.

Afin de fournir aux experts de domaine un outil leur permettant de
traiter de gros volumes de données, nous avons optimisé le temps de
traitement de l'ensemble des étapes de traitement des
documents. Les performances d'exécution du système,
testé sur un corpus de 8 500 documents, sont les suivantes~:

\begin{itemize}
\item Annotation des entités temporelles~: 8\,196 secondes,
\item Annotation des entités Agrovoc~: 1\,606 secondes,
\item Recherche des concepts et concepts liés
d'Agrovoc (Ressource Agrovoc \textit{offline})~: 36 secondes (en
utilisant le web service Agrovoc, ce temps peut être augmenté de 3
à 5 secondes par corpus),
\item Annotation des entités spatiales (français et anglais)~:
4\,940 secondes,
\item Génération vers le format choisi pour créer les index (JSON)~: 55 secondes.
\end{itemize}
Le processus prend un temps total de 16\,105 secondes, soit 1.9 secondes par document,
ce qui est très encourageant.

Actuellement nous travaillons à la mise en place du moteur de
recherche géographique permettant aux experts de domaines
d'analyser les corpus indexés. Dans ce sens, les
données indexées et validées sont intégrées dans une base de
données documentaire. Ce travail doit permettre,
d'une part, l'analyse des données et,
d'autre part, la recherche des publications portant sur
un même terrain et/ou une même période et/ou une même
discipline ou un sous-domaine à l'aide
d'un démonstrateur Web de recherche
d'information géographique.

\section{Conclusion}

Dans cet article, nous avons décrit la
démarche employée pour traiter un corpus de documents scientifiques
relatif au changement climatique, issus de trois organismes
différents. La question de la normalisation des données traitées
s'est évidemment posée. Nous avons ainsi
développé des algorithmes et un modèle de données unifié.
Actuellement, l'ensemble du corpus est indexé et au
format JSON. Nous travaillons actuellement à
l'enrichissement de la chaîne de marquage des
entités temporelles pour intégrer la solution BioTex ainsi que sur
l'extension des évaluations du marquage des entités
marquées (spatiales, temporelles et thématiques) sur des corpus
volumineux.

Nous travaillons également à
l'intégration des données dans le moteur de
recherche d'information multi-dimensionnelle
ElasticSearch\footnote{\url{https://www.elastic.co/fr/}} et sur
l'évaluation des données annotées à partir de
cas d'usages définis avec les experts géographes
membres du projet. Ce type de projet interdisciplinaire requiert une
gestion de projet particulière notamment sur la partie
{\guillemotleft}Exploration des publications et analyse des fronts de
recherche~». Comment faire parler ces données
constitue notre défi actuel.

\bibliographystyle{apalike-fr}
\bibliography{biblio}
% \bibliographystyle{iso690-fr}
% \bibliography{bj-bibtex}

\end{document}